\documentclass[12pt]{article}
\usepackage{epsf}
\usepackage{amsmath}
\usepackage{graphics}
\usepackage{graphicx}

\renewcommand{\thefootnote}{\#\arabic{footnote}}

\newcommand{\gtrsim}{ \mathop{}_{\textstyle \sim}^{\textstyle >} }
\newcommand{\lesssim}{ \mathop{}_{\textstyle \sim}^{\textstyle <} }
\newcommand{\gsim}{ \mathop{}_{\textstyle \sim}^{\textstyle >} }
\newcommand{\lsim}{ \mathop{}_{\textstyle \sim}^{\textstyle <} }
\newcommand{\ms}{M_{\odot}}

\begin{document}

\setcounter{footnote}{0}
\begin{titlepage}

\begin{center}

\hfill hep-ph/0608126\\
\hfill August 2006\\
\hfill KU-TP005\\

\vskip .5in

{\Large \bf
Constraining SuperWIMPy and Warm Subhalos \\
with Future Submillilensing
}

\vskip .45in

{\large
Junji Hisano$^1$, Kaiki Taro Inoue$^2$, 
and Tomo Takahashi$^3$
}

\vskip .45in

{\em
$^1$ Institute for Cosmic Ray Research, \\
University of Tokyo, 
Kashiwa 277-8582, Japan\\
$^2$ Department of Science and Engineering, Kinki University, \\
Higashi-Osaka, 577-8502, Japan  \\
$^3$ 
Department of Physics, Saga University, Saga, 840-8502, Japan
}

\end{center}

\vskip .4in

\begin{abstract}

  We propose to observe QSO-galaxy strong lens systems to give a new
  constraint on the damping scale of the initial fluctuations. We find
  that the future observation of submilliarc scale astrometric shifts
  of the multiple lensed images of QSOs would find $\sim 10^{(3-9)}
  M_{\odot}$ subhalos inside the macrolens halo.  The superweakly
  interacting massive particles (superWIMPs) produced from a WIMP
  decay and the warm dark matter (WDM) particles that predict a
  comoving damping scale larger than $\sim 2~\textrm{kpc} $ can be
  constrained if $\sim 10^3 M_{\odot}$ subhalos are detected.

\vspace{1cm}
\end{abstract}

\end{titlepage}

\renewcommand{\thepage}{\arabic{page}}
\setcounter{page}{1}
\renewcommand{\thefootnote}{\#\arabic{footnote}}
\renewcommand{\theequation}{\thesection.\arabic{equation}}

\section{Introduction}
\setcounter{equation}{0}

There has been mounting evidence that most of the matter in the
universe is not luminous but dark. Current observations such as the
cosmic microwave background (CMB) suggest that the dark matter (DM)
makes up about 25 \% of the universe \cite{Spergel:2006hy}. The dark
matter is usually assumed to be a collisionless cold component
(non-relativistic at the time of freeze-out), called as the cold dark
matter (CDM). Weakly-interacting massive particles (WIMPs), such as
the lightest neutralino in the supersymmetric standard model (SUSY SM)
\cite{susydm}, and the lightest Kaluza-Klein particle in the universal
extra dimension (UED) \cite{kkdm}, are the popular CDM candidates. The
predicted thermal relic abundances and the large-scale structure
($\gsim 1$ Mpc) are in good agreement with the observed values.

However, the recent high-resolution $N$-body simulations on the
CDM-based structure formation revealed various discrepancies on
smaller scales ($\lsim 1$ Mpc). The first one is so-called the
``missing satellite problem'' \cite{dwerf}: the $N$-body simulations
of the CDM particles predict significantly more virialized dark
objects with mass $M\lsim 10^{9}\ms$ (or subhalos) in the galaxy-sized
halos with $M\sim 10^{12} \ms$ than those observed around the Milky
Way.  The other one is called the ``cusp problem'' \cite{core}: the
CDM-based models also predict a cuspy profile for mass density
distributions for the CDM halos \cite{nfw} although the measurements
of the rotation curves imply the presence of cores in the centers of
the halo.

Although these discrepancies may be circumvented by some baryonic
processes \cite{UV}, it may be worthwhile to consider the other kinds
of DM particles with different clustering properties.  For instance,
the superweakly interacting massive particles (superWIMPs)
\cite{superwimp} or the warm dark matter (WDM) \cite{warm} particles
can have large velocity dispersion at the epoch of radiation-matter
equality.  If the DM consists of the superWIMPs or the WDM particles,
the number of less massive $\lesssim 10^9\ms$ subhalos is
significantly reduced, and the cusp formation is also suppressed
because the primordial fluctuations at the small scales ($\lesssim 1$
Mpc) are damped.

In this note, we consider the possibilities of probing such subhalos
with $M\lsim 10^{9}\ms$ via strong lensing in which the image
separations are on submilliarcsecond scales, called as
``submillilensing''.  Recently it has been pointed out that the future
submillilensing observations of multiply-imaged QSO-galaxy lens
systems can directly probe the mass scale of subhalos in the parent
galaxy halo \cite{inoue-chiba2003}.  The submillilensing observation
might resolve whether above problems originate from various baryonic
contrivances or the nature of the DM particles.  In section 2, we
begin with the discussion of submillilensing and consider the
possibility of detecting the small-mass subhalos via submillilensing
in the next decade. In section 3, we study its implications to the
superWIMP and the WDM scenarios.  Section 4 is devoted to Conclusions
and Discussion.

\section{Submillilensing}

In what follows, we discuss the possibilities of direct detection of
subhalos with a mass of $M\gsim 10^{3} M_{\odot}$ via substructure
lensing which is defined as lensing by $M\lsim 10^9 M_{\odot}$
subhalos that perturb a ``simple'' strong lensing by $\sim 10^{12}
M_{\odot}$ parent galaxy halo.  To date, about $10$ quadruply-imaged
gravitational lenses with flux ratio ``anomalies'' have been detected
\cite{anomaly1,anomaly2,anomaly3}.  Here ``anomaly'' refers to an
observed image flux ratio that does not agree with the ratio predicted
by standard macrolens models with a smooth gravitational potential.
From the radio and the mid-infrared observation, some of those lens
systems are found to be consistent with the model in which the
macrolens with a smooth potential is perturbed by subhalos
\cite{anomaly1, anomaly2, anomaly3, anomaly4,
  anomaly5}. Unfortunately, the mass scale of the subhalos has not
been yet determined well.
 
First, we theoretically estimate the mass range of the subhalos that
can perturb the fluxes of the multiple images produced by the parent
galaxy halo with a mass of $\sim 10^{12} M_{\odot}$.  In what follows,
for simplicity, we assume that the low-mass subhalos are described by
tidally-cut singular isothermal spheres (SISs) with a mass function
$dn/dm\propto m^{-2+\epsilon}$, where $0<\epsilon \ll 1$. At a
distance $r$ from the center, an SIS with a one-dimensional velocity
dispersion $\sigma_{\textrm{SIS}}$ has a density profile $\rho(r)
\propto \sigma_{\textrm{SIS}}^2/r^2$ \cite{gd}.  If we further assume
that the parent galaxy halo is also described by an SIS with a
one-dimensional velocity dispersion $\sigma$, then the tidal radius
$r_t$ of the SIS subhalo with $\sigma_{\textrm{SIS}}$ at a distance
$r$ from the center of the parent halo is approximately given by
$\sigma_{\textrm{SIS}}/r_t \approx \sigma/r$, which yields the mass of
the tidally-cut SIS as $m \propto r \sigma_{\textrm{SIS}}^3$.  The
effects of deviation from the assumption made here will be discussed
later.

The lensing cross section by a subhalo is proportional to the square
of the Einstein angular radius $\theta_E$, and the radius for the SIS
is proportional to $\sigma^2_{\textrm{SIS}}$ \cite{gd}.  Using the
equations for the mass function and the tidal radius, the substructure
lensing cross section per logarithmic mass interval is given by
\begin{equation}
d\tau \propto 
\theta_E^2 dn \propto \sigma_{\textrm{SIS}}^4 dn \sim
m^{1/3+\epsilon} d(\ln m)
\end{equation}
where we have assumed that the distance $r$ of the subhalo to the halo
center is approximately equivalent to the Einstein radius of the
parent halo $r_E$. This approximation is verified because the
substructure lensing cross section depends weakly on the distance to
the center $r$ as $\propto r^{-4/3}$.  The contribution from the
distant halos in the line of sight can boost the cross section by a
factor \cite{inoue-chiba1}.  Thus, the contribution of massive
subhalos is significant in the substructure lensing as long as the
mass function satisfies $\epsilon >-1/3$.  Assuming that the mass
function with $0<\epsilon<0.2$, observed for the massive scales
$M\gsim 10^9 M_{\odot}$ \cite{springel2001}, is applicable to the mass
scale $10^3 M_{\odot}\lsim M\lsim 10^9 M_{\odot} $, the lensing
probability by an intermediate-mass subhalo $\sim 10^3 M_{\odot}$ is
reduced by $\sim 10^{-(3-4)}$ in comparison with the probability by a
massive subhalo $\sim 10^9 M_{\odot}$.

If the subhalos have a different mass profile such as the NFW profile
\cite{nfw}, the contribution of less massive subhalos can be further
reduced.  Because the ratio between the Einstein radius of the
perturber and the tidal radius at a fixed distance $r$ decreases as
the subhalo mass decreases, the lensing cross section for a less
massive subhalo is susceptible to the inner less cuspy profile. Thus,
we expect that the contribution from $M\lsim 10^{3} M_{\odot}$
subhalos is negligible in alternating the amplitude of the flux of
multiple images.

Next, we explore the possibility of measuring the mass scale of these
subhalos from astrometric shifts of the multiply lensed images.  For
SISs, the order of the magnitude of the astrometric shifts is
typically given by the size of the angular Einstein radius $\theta_E$
\cite{inoue-chiba2}.  For an SIS with one-dimensional velocity
dispersion $\sigma_{\textrm{SIS}}$ at $\sim$Gpc, $\theta_E$ is
approximately given by
\begin{equation}
\theta_E \sim 10  \biggl (  
\frac {  \sigma_{\textrm{SIS}} } {20~\textrm{km~s}^{-1}  }     
\biggr )^2 \textrm{mas}.
\end{equation}
Thus, observation with angular resolution of submilliarcsecond scales
($\sim 0.01$mas), which will be achieved in the next generation
satellite VLBI mission such as the VSOP2 \cite{vsop2}, can reveal
subhalos with one-dimensional velocity dispersion
$\sigma_{\textrm{SIS}} > 0.6$ km~s$^{-1}$. It corresponds to $M\gsim
10^3 M_{\odot}$ at the distance equal to the Einstein radius $r=r_E$
of the macrolens, assuming that the velocity dispersion of the parent
halo is $\sigma \sim 200~\textrm{km~s}^{-1}$.

From the astrometric shifts of the multiple extended images perturbed
locally by a subhalo with respect to an unperturbed macrolensed image,
we can break the degeneracy between the subhalo mass and the distance
in the line of sight to the images if resolved at scale of an Einstein
radius of the perturber \cite{inoue-chiba1, inoue-chiba2}.  This is of
great importance because otherwise we cannot determine whether the
flux ratio anomaly is caused by more massive intergalactic halos in
the line of sight or by less massive subhalos within the macrolens
halo.  Even if the density profile of the perturber is shallower than
an SIS, we can make a distinction between models with different
density profiles from astrometric shifts of the surface brightness
profile within the source \cite{inoue-chiba2}.  A direct detection of
less massive $10^{3} M_{\odot}\lsim M\lsim 10^{9}M_{\odot}$ subhalos
within the parent halo will give a stringent constraint on the
superWIMP and the WDM scenarios, which will be discussed in the next
section.

\section{Implications to SuperWIMP and WDM Scenarios}
There exist many well motivated models for superWIMPs and WDM from
particle physics. The natural candidates for the superWIMPs are
gravitino and axino, that are superpartners of graviton and axion,
respectively \cite{superwimp}. The right-handed sneutrino is also the
candidate when neutrino masses are Dirac-type \cite{asaka}.  Others
are Kaluza-Klein graviton and axion states in the UED
\cite{superwimp}.  As for the WDM, a light gravitino and sterile
neutrinos have been discussed as such candidates.  Thus, it is
interesting to study the feasibility of probing or constraining those
models in the near future experiments.

In the following, we refer ``superWIMPs'' as the particles whose
interactions are weaker than the weak interaction, such as the
gravitino which couples to other particles only gravitationally.
SuperWIMPs are also assumed to be produced by the decays of heavier
particles whose interactions are weak ({\it e.g.}
WIMPs)\footnote{
  Usually conventional WDM particles do not satisfy both of these
  properties.  In this sense, superWIMPs can be distinguished from
  conventional WDM particles.
}.

In the CDM-based structure formation models, the structure of the
universe forms in a hierarchical manner. Protohalos, which are the
first virialized objects, appear first after the mass density
fluctuation becomes nonlinear, and larger objects form successively
via their merger.  In the ordinary WIMP models, the comoving damping
scale of the power spectrum is typically $(0.01-10)$pc, depending on
the Hubble radius at the kinetic decoupling temperature
\cite{Loeb:2005pm}, and the protohalo mass is $(10^{-12}-10^{-4})
M_{\odot}$ \cite{Profumo:2006bv}.

In the superWIMP or WDM scenarios, the primordial fluctuations on even
larger scales can be damped because they have large velocity
dispersion at the decoupling.  Therefore, the protohalos become
massive, and the total amount of subhalo mass inside the parent halo
can be significantly reduced.

When the comoving damping scale
  of the DM power spectrum is $R_{\rm cut}$, the protohalo mass is
  roughly given by
\begin{eqnarray}
M_{\rm cut} &\simeq& 1\times 10^{11}M_{\odot}\left(\frac{R_{\rm cut}}{1{\rm Mpc}}\right)^3.
\end{eqnarray}
Here, the present matter density $\Omega_{m,0}=0.24$ and the Hubble
constant $h=0.73$ are assumed \cite{Spergel:2006hy}.  In the previous
section, we showed that the sensitivity of direct detection of
subhalos inside the parent halo with submillilensing may reach
intermediate mass $\sim 10^{3}M_{\odot}$ scales.  Thus, when $R_{\rm
  cut}\gsim 2 $ kpc, the future submillilensing experiments may
directly detect such protohalos.

First, we discuss the damping scale in the superWIMP scenario and
compare it with the sensitivities of future submillilensing. In the
scenario, some WIMPs freeze out from the thermal equilibrium as the
usual WIMP DM models, and superWIMPs are non-thermally produced from
the WIMP decay, since the superWIMPs interact superweakly with the
thermal bath.  The scenario retains the property of the ordinary WIMPs
that the observed relic density is naturally achieved.  Furthermore,
they are produced by decay of some long-lived massive particle $X$.
Since they can have large velocities at the production epoch, the
free-streaming damps the small-scale inhomogeneities.  The superWIMP
scenario is a possible explanation for the small-scale structure
\cite{superwimpsd1,superwimpsd2}.

The comoving damping scale is typically given by the free-streaming
length of the superWIMP, $R_{\rm fs}$, at the matter-radiation
equality $t_{\rm eq}$. When the long-lived particle has lifetime
$\tau_X$ and the superWIMP is produced with three-momentum normalized
by its mass $u(=p/m)$, the comoving free-streaming scale $R_{\rm fs}$
is given by
\begin{eqnarray}
R_{\rm fs}
&=&
\frac{2 t_{\rm eq}}{a(t_{\rm eq})}
u_{\rm eq}
\left[
\log\left(\frac{1}{u_{\rm eq}}+\sqrt{1+\frac{1}{u_{\rm eq}^2}}\right)
-
\log\left(\frac{1}u+\sqrt{1+\frac{1}{u}}\right)
\right]
\end{eqnarray}
where $u_{\rm eq}=(a(\tau_X)/a(t_{\rm eq})) u$ and $a(t)$ is the scale
factor as a function of $t$.

When the superWIMPs are produced from decay of electrically-charged
particles\footnote{
  {It is recently pointed out in Ref.~\cite{Pospelov:2006sc} that
   long-lived charged particles ($\tau_X \gsim 10^3$sec) lead to
    overproduction of $^6$Li in the Big Bang Nucleosynthesis due to
    the catalytic enhancement of the production. On the other hand, it
    is argued in Ref.~\cite{Kohri:2006cn} that there are a lot of
    ambiguities in the derivation.}
}, the small-scale power may be further suppressed
\cite{Sigurdson:2003vy}.  The charged particles are coupled to the
photon-baryon fluid which oscillates on subhorizon scale. If the scale
in question enters the horizon before the decay, the density
fluctuation of such DMs cannot grow but oscillates, which gives a rise
to the suppression on small scales. The damping scale is typically
given by the Hubble radius at the decay time, $ H^{-1}(\tau_X)$,
\begin{eqnarray}
R_{\rm ch}
&=&\frac{1}{a(\tau_X)} H^{-1}(\tau_X).
\end{eqnarray}

In Fig.~1 we show the damping scales $R_{\rm cut}$ of the charged and
neutral $X$ cases as functions of $\tau_X$ and $u$.  When $X$ is
neutral, the damping scale is determined by the free streaming scale
$R_{\rm fs}$. Longer lifetime $\tau_X$ implies a larger free streaming
scale, since the velocity dispersion at the radiation-matter equality
time, which is proportional to $a(\tau_X)/a(t_{\rm{eq}})$, becomes
larger.  When $X$ is charged, the damping scale $R_{\rm cut}$
corresponds to the larger one of $R_{\rm fs}$ and $R_{\rm ch}$.  For
$u\ll 1$, $R_{\rm cut}$ is given by $R_{\rm ch}$, since $R_{\rm fs}$
is $\sim u \times R_{\rm ch}$.  The hatched region with damping scales
larger than $\sim 1$ Mpc is constrained from Lyman alpha clouds
\cite{lyman}. The future submillilensing experiments may cover regions
above the bold lines, which correspond to $R_{\rm cut}\gsim 2$
kpc. When the $X$ lifetime is longer than $\sim 1$ sec and the
superWIMPs are produced with relativistic momentum $(u\gsim 1)$, the
damping scale is larger than $\sim 1$ kpc, which may be constrained by
future observation if the subhalos with $M\sim 10^3 M_{\odot}$ were
discovered.  When $X$ is charged, the region with $\tau_X\gsim 400$
sec may be also covered even in the small $u$ cases.

\begin{figure}[t]
\centerline{\epsfxsize = 0.6\textwidth \epsffile{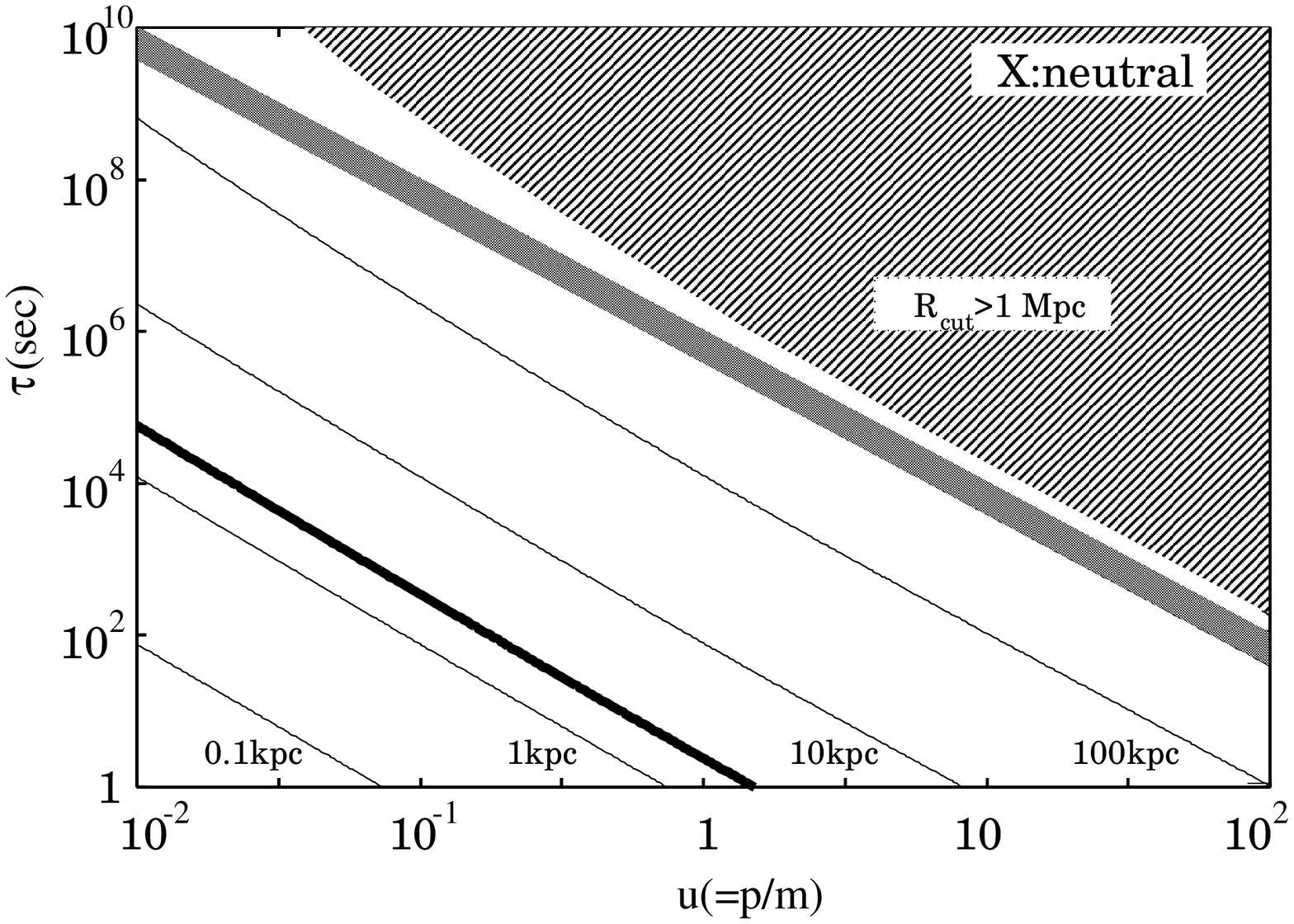} }
\centerline{\epsfxsize = 0.6\textwidth \epsffile{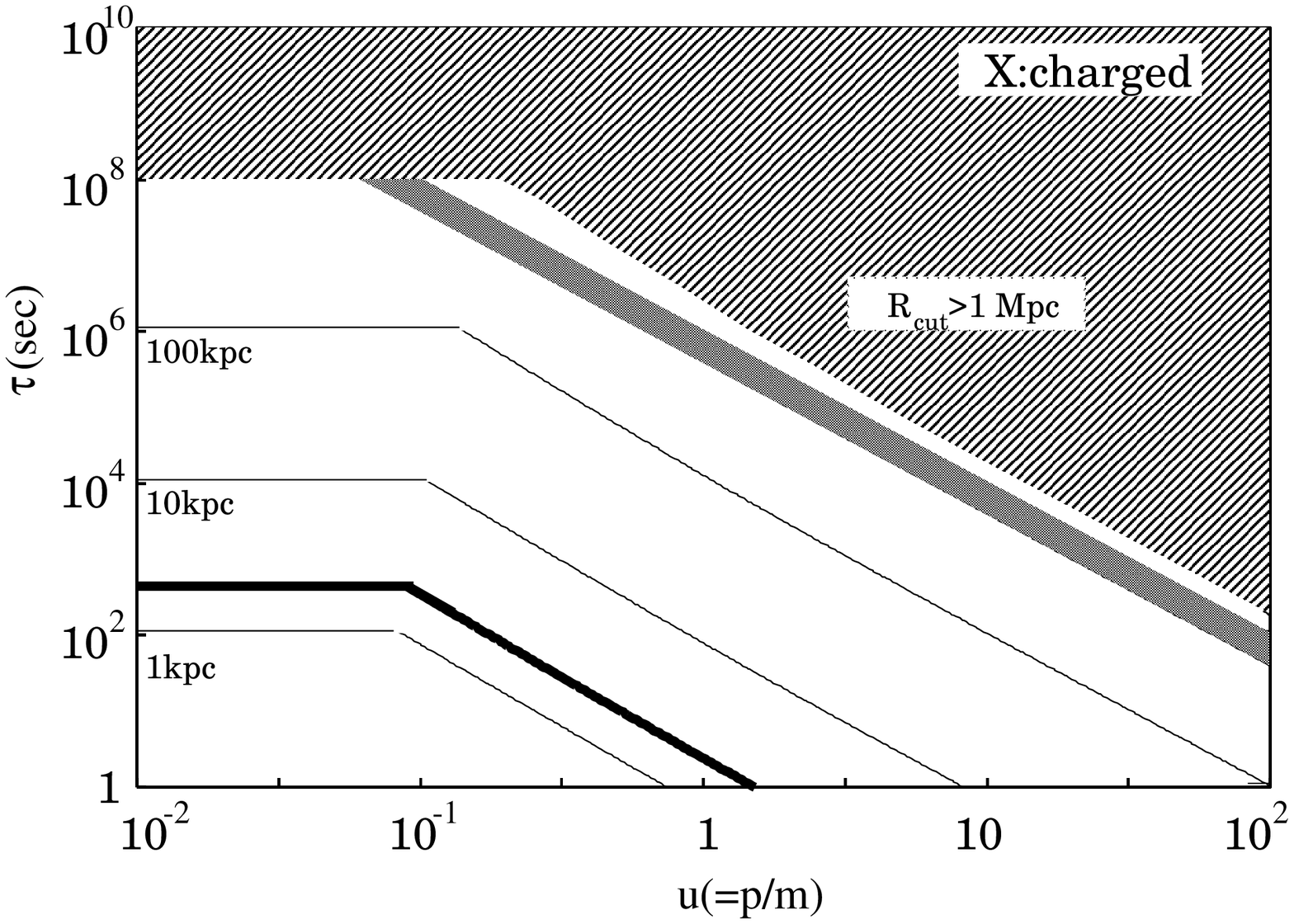} }
\caption{
  Damping scales of the DM power spectrum as functions of $\tau_X$ and
  $u$ in the superWIMP scenarios. The upper figure is for the neutral
  $X$ case, and the lower one is for the charged $X$ one.  The
  (hatched) region with damping scale larger than 1 Mpc is constrained
  from Lyman alpha clouds.  The gray region is favored to solve the
  cusp problem. The future submillilensing observations may cover the
  region above the bold lines which correspond to $R_{\rm cut}\gsim 2$
  kpc.}
\label{fig1}
\end{figure}

Following Ref.~\cite{superwimpsd1}, we indicated the parameter region
which is suitable to solve the discrepancies in small-scale structure
in Fig.~1. The region with $R_{\rm cut}\simeq (0.4-1.0)$ Mpc is
suitable to solve the ``missing satellite problem'', while the gray
region with $1\lsim u^{-3}(\tau_X/10^6 {\rm sec})^{3/2}\lsim 4$ is
favored to solve the ``cusp problem'', which requires a large DM
velocity dispersion \cite{Hogan:2000bv}. As we can see in Fig.~1, the
model parameters corresponding to these regions can be well
constrained by the future submillilensing observation.

For the illustrative purpose, we consider a superWIMP gravitino model,
in which the gravitinos are produced by slepton or sneutrino decay.
Sleptons or sneutrinos are assumed to be the lightest SUSY particles
in the SUSY SM\footnote{
  The lightest neutralino is also one of the candidates for the
  lightest SUSY particle in the SUSY SM. However, the decay produces
  hadronic shower to spoil the BBN if it is not photino-like. The
  long-lived slepton and sneutrino are not strongly constrained by the
  BBN, since their hadronic branching ratios are small. See
  Ref.~\cite{Feng:2004zu} for the constraints on the superWIMP
  gravitino model.
  }.  The slepton and sneutrino lifetimes
are given by \cite{Feng:2003uy}
\begin{eqnarray}
\tau_X &=&48\pi M_\star^2\frac{m^2}{m_X^5}
\left(1-\frac{m^2}{m_X^2}\right)^{-1}
\end{eqnarray}
where $M_\star = 2.4\times 10^{18}$ GeV, $m$ is the gravitino mass,
and $m_X$ is the mass of the parent WIMP $X$.  In Fig.~2, $R_{\rm fs}$
(solid lines) is shown as a function of $m$ and the mass difference
between the gravitino and the parent particle ($\Delta
m=m_X-m$). Dashed lines indicate the damping scale for which $R_{\rm
  ch}>R_{\rm fs}$.  The gray region which corresponds to explain the
``cusp problem'' \cite{superwimpsd1} may be excluded when the future
submillilensing experiments find subhalos with mass smaller than $\sim
1\times 10^{(7-8)} M_\odot$.  It is known that the lifetime and the
mass of the long-lived particles whose decay produces the superWIMPs,
are constrained by the Big Bang Nucleosynthesis (BBN) and by the CMB
Planckian spectrum, depending on the decay channels. When the hadronic
modes are dominant in the decay, the BBN constrains the lifetime to be
shorter than $\sim 1$ sec \cite{Kawasaki:2004yh}.  Even if the
hadronic shower is suppressed, the electromagnetic energy injection
from the long-lived particle decay to the thermal bath is also
constrained from the BBN while the constraint is weaker. In addition
to it, the CMB Planckian spectrum also gives a constraint when the
lifetime is longer than $\sim 10^6$ sec \cite{Feng:2003uy}.

The future submillilensing experiments are complementary to those
constraints, since the damping scale $R_{\rm cut}$ is independent of
the decay channels. For example, when the superWIMP gravitinos are
produced from sneutrino decay, the energy injection to thermal bath is
very tiny so that the constraints from the BBN and the CMB Planckian
spectrum are very weak. Even in such a case, the submillilensing will
still constrain the model.

\begin{figure}[t]
\centerline{\epsfxsize = 0.6\textwidth \epsffile{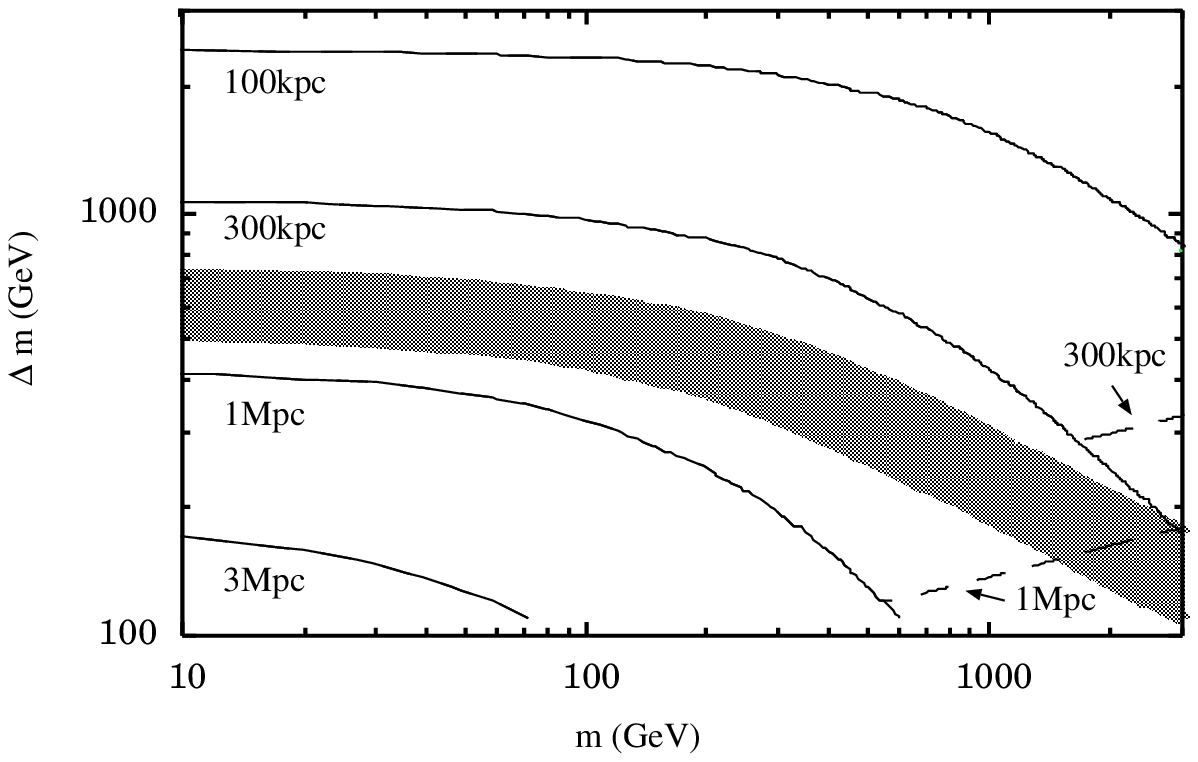} }
\caption{Damping scales for the gravitino superWIMPs produced by
  slepton or sneutrino decay.  $R_{\rm fs}$(solid lines), $R_{\rm
    ch}$(dashed lines) are plotted as functions of $m$ and $\Delta
  m=m_X-m$, respectively.  $R_{\rm ch}$ is shown only for the case
  $R_{\rm ch}>R_{\rm fs}$.  The gray region is favored to solve the
  cusp problem.  }
\label{fig2}
\end{figure}

Next, we discuss the damping scale in the WDM models.  Among many WDM
candidates discussed so far, the light gravitino with mass of order
from $10^{-6}~{\rm eV}$ up to $1~{\rm keV}$, which is likely to be the
lightest SUSY particle in gauge-mediated models of supersymmetry
breaking, is one of the well-motivated models from the viewpoint of
particle physics \cite{Pagels:1981ke,Bond:1982uy}.  Such gravitinos
are in thermal equilibrium at early times but decouples when the
degrees of freedom $g_*(T_D)$ is $\mathcal{O} ( 100 )$ where $T_D$ is
the decoupling temperature. Because the decoupling temperature of such
species is higher than that of (active) massive neutrinos which play
roles of hot dark matter, their velocity dispersion is not so large
compared to that of massive neutrinos but non-negligible at the time
of structure formation. Thus, they can act as the WDM.  The comoving
damping scale for the free-streaming for such gravitinos or any other
thermal relic can be written as
\begin{equation}
R_{\rm fs} 
\sim 0.84 {\rm Mpc}  \left(\frac{g_*(T_D)}{10.75}\right)^{1/3} 
\left( \frac{1 {\rm keV}}{m_{\rm WDM}} \right) 
\left( \frac{\langle p/T\rangle }{3.15} \right),
\label{eq:damping_WDM}
\end{equation}
where $\langle p/T\rangle$ is the mean momentum over the temperature.
For thermally decoupled species, this factor gives almost unity, {\it
  i.e.}, $\langle p/T\rangle/3.15 \sim 1 $.  The requirement from
Lyman alpha clouds, $R_{\rm fs} \lsim 1 {\rm Mpc}$, implies $m_{\rm
  WDM} \gtrsim 1 {\rm keV}$. On the other hand, the energy density of
WDM can be written as $ \Omega_{\rm DM} h^2 = ( m_{\rm WDM} / 94 {\rm
  eV}) ( 10.75 / g_*(T_D) )$. Assuming $\Omega_{\rm DM}h^2 \sim 0.10$,
the mass of WDM should be $ m_{\rm WDM} \sim 0.1$ keV even for
$g_*(T_D)\sim 100$. We need to introduce more extra degree of freedom
around $T_D$ as $g_*(T_D)\sim \mathcal{O}(10^3)$.  When the constraint
on $R_{\rm fs}$ is improved to be $\lesssim 1$kpc, the lower bound for
the mass can reach $m_{\rm WDM} \sim 1$ MeV for $g_*(T_D)\sim 100$.
Thus, the WDM scenario may face further difficulties if future
submillilensing experiments would find small-mass subhalos.

Another well-motivated candidate for WDM is the sterile neutrinos
\cite{Dodelson:1993je}.  Because they directly couple to the active
neutrinos alone, they can be produced via neutrino oscillation.
Although the evaluation of their energy density requires a numerical
integration of the Boltzmann equation, some useful fitting formulae
are available.  The present energy density of sterile neutrinos can be
written as \cite{Abazajian:2005xn}
\begin{equation}
\Omega_{\nu_s} h^2 \sim 
0.3 \left( \frac{\sin^2 2 \theta }{10^{-10}} \right) 
\left( \frac{m_s}{100 {\rm keV}} \right)^2,
\end{equation}
where $\theta$ is the mixing angle between the active neutrinos and 
sterile neutrinos and $m_s$ is the mass of sterile neutrinos.
The temperature at the time when the production is most efficient is 
\cite{Abazajian:2005xn,Abazajian:2005gj}
\begin{equation}
T_{\rm peak} \sim 130 {\rm MeV} \left( \frac{m_s}{3 {\rm keV}} \right)^{1/3}.
\end{equation}

The sterile neutrinos can damp the small-scale inhomogeneities via the
free streaming in the same manner as the thermally decoupled WDM
particles do.  The free-streaming scale for a sterile neutrino WDM can
also be obtained using Eq.~(\ref{eq:damping_WDM}) with different
values for $\langle p/T\rangle$ from that of the thermally decoupled
ones.  Because the sterile neutrinos are not in the thermal bath at
early times, their distribution function deviates from that of a
thermal one and the above factor can be $\langle p/T\rangle/3.15 \sim
0.9 $ for the standard production mechanism \cite{Abazajian:2006yn}.
Although there are some differences between the thermally decoupled
WDM and the sterile neutrino WDM, they give the same predictions for
the damping of matter power spectrum by identifying their masses as
\cite{Colombi:1995ze,Viel:2005qj}
\begin{equation}
m_{s} 
= 4.71 {\rm keV} 
\left( \frac{m_{\rm thermal} }{1 {\rm keV} } \right)^{4/3}
\left( \frac{0.10}{\Omega_{\rm DM}h^2} \right)^{1/3},
\end{equation}
where
  $m_{\rm thermal}$ is the mass of thermally decoupled relics such as
  a light gravitino, which is denoted as $m_{\rm WDM}$ in
  Eq.~(\ref{eq:damping_WDM}).
Because the shape of matter power spectrum is determined by the ratio
$m/T$ and the density parameter $\Omega_{\rm DM}h^2$.  
Accordingly, the constraint on the
sterile neutrino mass bound would be different from that on the
thermally decoupled WDM mass bound\footnote
{The properties of the sterile neutrinos can also be constrained from
  the measurement of the X-ray flux, since they can contribute to the
  X-ray flux due to radiative decay \cite{Xray}. See also
  Refs.~\cite{Abazajian:2006yn,Viel:2006kd}.}.
Using the above formula and assuming the damping scale as $R_{\rm cut}
\lesssim 1$ kpc which can be reached by future submillilensing
experiments, we can expect that the mass of sterile neutrinos can be
constrained to be $m_s \gtrsim 40$ MeV, which will be in conflict even
with the current constraint $m_s \lesssim 10$ keV
\cite{Abazajian:2006yn,Xray,Viel:2006kd}.  Thus, we can obtain much
more insight on the robustness of these scenarios from the future
submillilensing experiments, as well as for the superWIMP scenarios.

Here some comments on the mixed dark matter scenario are in order.  It
is possible that, for example, gravitinos are produced not only from
the decay of the next-lightest supersymmetric particles (NLSP) but
also from thermal plasma. In this case, the present DM is composed of
CDM and superWIMP. In such mixed DM scenarios, the damping of the
small-scale structure is less significant in comparison with the
models in which the superWIMPs make up all of the DM.  To what extent
the amplitude at small scales can be reduced depends on the ratio of
the energy density of CDM and superWIMP DM.  Detailed discussion on
this issue is beyond the scope of this letter.  Some discussions on
the matter power spectrum in such mixed models can be found in
Refs.~\cite{Profumo:2004qt} and
\cite{Kaplinghat:2005sy}. \cite{Profumo:2004qt} analyzed models in
which the superWIMPs are produced from the charged particle decay
while \cite{Profumo:2004qt} considered models in which the superWIMPs
are produced from the neutral particle decay.

\section{Conclusions and Discussion}
We have shown that future observation of multiply-imaged QSO-galaxy
lens systems with a high angular resolution $\sim 0.01$ mas will prove
the small-scale clustering properties of the DM halos down to $\sim
10^3 \ms$.  The presence of $\sim 10^{3}M_{\odot}$ subhalos implies
the comoving damping scale of the primordial fluctuations $\sim 2$ kpc
assuming that the observed subhalos retain the original mass during
the merger process.  The superWIMP and the warm DM scenarios that
predict a larger damping scale $(0.002-1)~\textrm{Mpc}$ in comparison
with $(0.01-10)$ pc in the ordinary WIMP scenarios would be
strongly constrained if the presence of such subhalos were proved.

In the superWIMP scenario, the superWIMP DM is produced from the
long-lived particle $X$ decay, and free-streaming of the superWIMPs
damps the small-scale inhomogeneities.  When the lifetime of $X$ is
longer than $\sim 1$sec, the damping scale is larger than $\sim 1$ kpc
unless the superWIMP and $X$ masses are degenerate and the superWIMPs
are non-relativistic at the production epoch. In addition, when $X$ is
a charged particle, it is coupled to the oscillating photon-baryon
fluid before the decay. Therefore, the small-scale inhomogeneities
inside the sound horizon cannot grow.  The damping scale becomes
larger than $\sim 1$ kpc$\times(\tau_X/100{\rm sec})^{1/2}$ in the
case of charged $X$. One of the natural superWIMP candidates is the
gravitino produced from the slepton or sneutrino. In this case, the
damping scale is typically larger than $10^{2}$ kpc. The future
submillilensing experiments, which cover the DM subhalos with mass
$\gsim 10^{3}M_{\odot}$, are important tests for probing such
superWIMP scenarios. For the WDM scenarios, such as the light
gravitino and the sterile neutrinos, the WDM mass would be further
constrained.

The survivability of the protohalos during the merger process is under
debate now. In the ordinary WIMP scenarios, it is claimed that most of
the earth-mass protohalos are stable against the tidal stripping
\cite{protohalo}.  It is also discussed whether those halos are
disrupted by interaction with stars \cite{protohalo2}. On the one
hand, subhalos that cross the galactic disk nearly perpendicularly or
that fall off to the center of the parent galaxy are strongly
disrupted by the tidal force, leading to a significant decrease in the
total mass. On the other hand, subhalos that orbit on the plane nearly
parallel to the disk or that reside in the low density region survive
more-or-less intact.

For simplicity, we have assumed that the observed mass scale of the
subhalo is equivalent to the lowerbound on the protohalo mass
scale. In practice, however, the observable mass using gravitational
lensing is limited to the one within a certain radius centered at the
line of sight.  As a result, there remains a certain ambiguity in
estimating the mass scale of the observed subhalo. Observation of
astrometric shifts of lensed QSO images with a substructure in the
surface brightness may help to reconstruct the subhalo mass density
profile, thereby reducing the ambiguity \cite{inoue-chiba2003}.

It has been argued that the superWIMP and WDM scenarios can resolve
the small-scale $\lesssim 1 ~ \textrm{Mpc}$ discrepancies, such as the
``missing satellite problem'' and the ``cusp problem'' if the damping
scale is as large as $(0.4-1.0)$ Mpc.  Therefore, future
submillilensing experiments will shed a new light on these small-scale
structure problems once the above ambiguities are removed.

\bigskip 
\noindent
\section*{Acknowledgment:} 
We would like to thank Masashi Chiba for useful comments.  This work
is supported in part by the Grant-in-Aid for Science Research,
Ministry of Education, Science and Culture, Japan (No.~18034002 and
15540255 for JH).

\end{document}